# One-step Patterning of Sub-wavelength Plasmonic Gratings in Metal-Polymer Composites

**Raghvendra P Chaudhary, Govind Ummethala, Arun Jaiswal, Suyog R Hawal, Sumit Saxena and Shobha Shukla***

*Nanostructures Engineering and Modeling Laboratory, Department of Metallurgical Engineering and Materials Science, Indian Institute of Technology Bombay, Mumbai, MH, India 400076*

*[sshukla@iitb.ac.in](mailto:sshukla@iitb.ac.in)

**Abstract:**

2D and 3D micro/nano fabrication based on two-photon polymerization (TPP) has emerged as a strong contender for additive manufacturing for wide variety of applications. In this manuscript we report one step patterning of structurally stable, subwavelength 2D and 3D gold nanostructures using femto-second laser by incorporating single photon photoinitiator only in pure and metal precursor doped polymers. The metal polymer composite nanostructures are written directly by in-situ reduction of gold precursor within the photoresist using femto-second laser irradiation. The photo-initiator triggers the reduction of gold precursor and induces simultaneous polymerization of the photoresist based on two-photon absorption phenomenon. Diffraction gratings with varied loading of gold precursors in photoresist have been fabricated and characterized by measuring their diffraction efficiencies in the infrared region. Minimum line width of 390 nm has been achieved for 5 wt% gold loaded polymers. Systematic studies of the effect of laser power, metal precursor loading etc. on the line-widths have also been performed. These investigations are expected to pave way for exploration of new material combinations for development of simple fabrication for fabrication of metal composite functional structures at significantly low cost and with high-throughput.

**Introduction:**

Two-dimensional (2D) and three-dimensional (3D) metallic micro/nano structures have been explored in various fields like MEMS/NEMS, nanophotonics, optoelectronics, and biomedical applications, to name a few [1],[2],[3],[4],[5]. 1D and 2D subwavelength period diffraction metallic gratings have been used for numerous applications in optoelectronics for pulse compression[6], enhancing solar cell efficiency[7], and for polarizing light [8],[9] etc. Fabrication of such gratings supporting plasmon resonance is not only complicated but also requires time consuming multistep processes. In this context two-photon lithography (TPL) is of particular interest and allows direct fabrication of sub-wavelength 2D/3D structures near the polymerization threshold based on photo-polymerizable material [10],[11].[12]'[13]'[14],[15]'[16]. Since TPL is restricted to polymeric materials [17], one of the current challenge is the direct writing of functional structures inside metal-polymer composite. Since metals do not form self-supporting micro/nano structures due to their weak mechanical strength[18], hence patterning of metals have become a bottleneck in verification of theoretical concepts relevant to plasmonics and metamaterials,[19],[20],[21],[22],[23],[19],[23],[24]. Metal loaded polymer composite structures have been fabricated in PVA & SU8 matrix using TPL previously[25],[26]. However PVA-gold composite structures are huge where as SU8 reduces gold precursor immediately after mixing thereby leading to complications.

Traditionally, TPL utilizes two-photon absorbing dyes for fabricating subwavelength structures. However, these two-photon dyes are expensive and hence limits the potential of this fabrication technique [4],[27]. Dyeless TPL fabrication using inexpensive, commercially used photoinitiators such as Irgacure, Lucirin TPOL, PC2506 etc. have been explored in the past. However, the resolution of these structures could be achieved up to few microns only due to small two photon absorption (TPA) coefficient [28],[29],[11],[30], [31]. However high radical quantum yield of 0.99 and better solubility in most of the resins[32] makes Lucirin TPOL a better choice for patterning finer structures using TPL.

Here we report fabrication of metallic sub-wavelength structures in polymer matrix using 20 wt% lucirin TPOL loading and without any dye molecule. Using this modified chemistry, we have fabricated highly stable 2D and 3D gold-polymer composite functional structures such as optical gratings with different periodicities. These were then characterized by calculating diffraction efficiency. The formation of gold-polymer composite was verified using secondary and back scattered micrographs obtained using scanning electron

microscopy(SEM). We have further systematically investigated the effect of laser power, metal precursor loading etc. on the line-widths. This advancement is expected to simplify fabrication of complex 2D & 3D pasmonic structures at significantly low cost and with high-throughput.

**Experimental Procedure:**

Gold precursor loaded two liquid resins; triacryclicmonomers, tris (2-hydroxy ethyl) isocyanurate triacrylate (SR368) and ethoxylated (6) trimethylolpropane triacrylate (SR499) [Sartomer] were subjected to femtosecond laser irradiation. SR499 reduces structural shrinkage, whereas SR368 confers hardness to the structure during the laser irradiation. The samples were prepared by mixing SR368 and SR499 in the ratio of 48:49 by weight percent (wt %). 20wt% of a photoinitiatorethyl-2,4,6-Trimethylbenzoylphenylphosphinate (Lucirin-TPOL from BASF] was added to this mixture. Different wt% of gold (III) chloride trihydrate ($HAuCl_4.3H_2O$) powder [Alfa Aesar] was loaded in the above solution. Finally a ~ 3μm thick film was spin coated uniformly on a cover glass. Prior to spin coating, the cover glass was silanized with 3-aminopropyl triethoxysilane to facilitate the adhesion of the mixed resin. Acrylate-coated cover glass was positioned on a XYZ piezo-stage (PI-nanopositioner E-725) of an inverted optical microscope [Olympus-IX81]. The samples were then irradiated with Ti-sapphire laser source (Coherent made, Chameleon Ultra I) at 800nm wavelength, 140fs pulse width and a repetition rate of 80MHz via acousto-optic modulator as shown in figure S1 (supporting information). A 100X objective lens with numerical aperture of 0.9 was used to focus the beam inside the resin. An EMCCD attached to microscope facilitates the real time monitoring of the fabrication process. The experimental setup was controlled using LabVIEW and various scan rates were employed to fabricate the gratings. The unexposed resin was washed off in dimethylformamide (DMF) after writing. Since the temperature rise in the focal volume is sufficient enough for annealing [33], no post baking of the sample was performed.

Several samples of gold-polymer composite lines with 1wt%, 5wt% and 10wt% gold loading were fabricated. Structural analysis was performed by measuring the line width of the grating in the micrographs obtained from scanning electron microscopy (SEM). Functional grating structures with 1.5μm, 2.5μm and 5μm periodicity were fabricated inside pure polymeric resin as well as in 1wt%, 5wt% and 10wt% gold precursor loaded polymeric matrix. Large area SEM micrograph (560μm x 280μm) of the gratings fabricated inside 5wt% gold loaded polymer is shown in figure 1(a). The periodicities of the gratings are 1.5μm, 2.5μm and 5μm for bottom to top gratings respectively. Figure 1(b) is the zoomed in SEM micrograph of the 1.5μm period grating showing finely spaced lines. The colour images in figure S2 are the optical images of the large area (840μm x 840 μm) gratings captured in reflection mode using smartphone at different angles. EDAX spectrum for the sample in figure 1(c) shows strong signal at around 2.30 keV, which is in good agreement with the standard data. The presence of the silicon, sodium and potassium peaks in EDAX spectrum is from the glass cover slip. Further, the presence of gold nanoparticles was confirmed using backscattered electron (BSE) micrographs shown in figure 1(d). Large size gold nanoparticles embedded in polymer have been marked by yellow dotted circle in inset of figure 1(d). The full scale backscattered electron imagesare shown in figure S3.

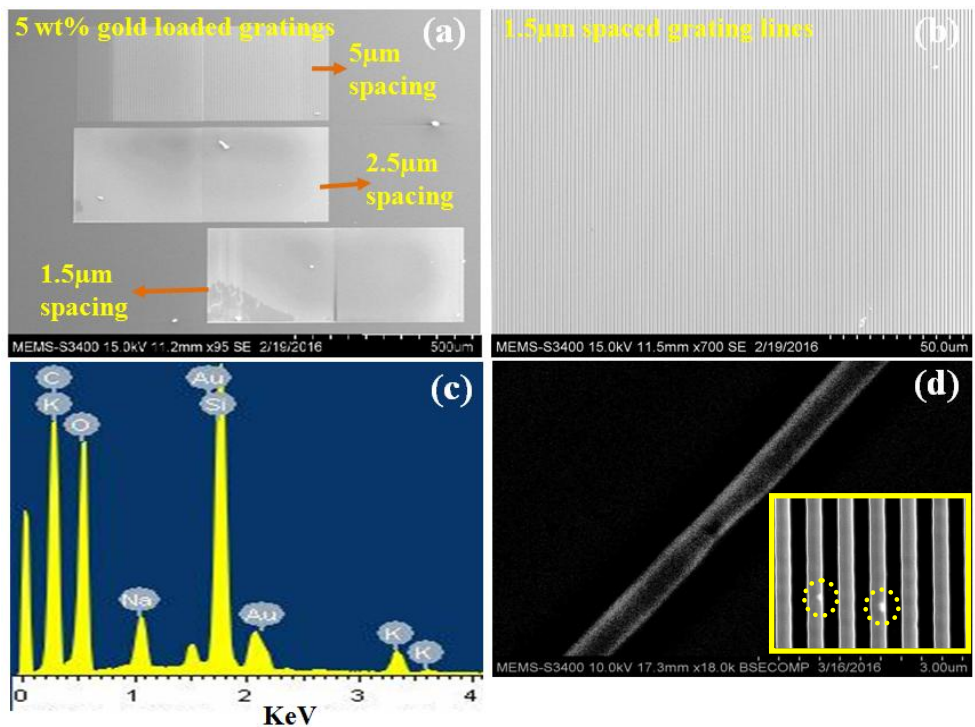

*Figure 1: a) Large area (560μm X 280μm) SEM micrographs of optical gratings fabricated in 5wt% gold loaded polymer matrix with periodicity of 1.5μm, 2.5μm and 5μm for bottom to top respectively. (b) Zoomed in image of 5wt% gold loaded grating showing the sturdy structure. (c) EDAX spectrum showing the presence of gold in the fabricated nanostructure. (d) Backscattered electron image showing the presence of gold naoparticles. Inset shows the secondary electron image of the lines with chunk of gold nanoparticle (marked with yellow circles).*

Diffraction patterns from the grating with periodicity 1.5μm, 2.5μm and 5μm fabricated inside 10wt% gold loaded polymer matrix (SEM images in figure 2(a-c))has been recorded using 712 nm laser at normal incidence and are shown in corresponding pannels respectively. Well-resolved diffraction patterns are evidence to the high quality of grating structures. Decreasing of line spacing shows an increase in the diffraction angle, hence following the Bragg equation of diffraction. The diffraction patterns obtained from a grating and 2D mesh structure are shown in figure S4.

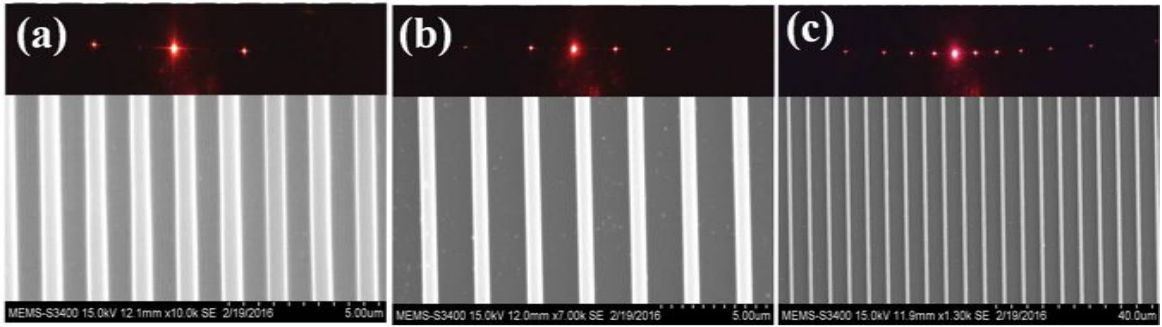

*Figure 2: Diffraction pattern of red laser (712 nm) from gratings with (a) 1.5μm, (b) 2.5μm and (c) 5μm periodicity fabricated inside 5wt% gold loaded polymer*

The diffraction efficiency of the grating structures fabricated in pure polymer, 0 wt%, 5 wt% and 10wt% gold loaded polymer matrix with constant periodicity of 1.5 μm has been compared in figure3. Tunable femtosecond laser source in the wavelength range 800nm-1000nm and constant laser power of 400 mW has been employed for these measurements. The diffraction efficiency of gold loaded grating structures has been found to be more than that of pure polymer and is maximum for 10 wt % gold loading. This effect could be attributed to the broader line-width of the undoped polymer where the spacing between the lines is smaller than the wavelength used hence the diffraction is weak. With increase in the gold concentration the line width decreases and the spacing between the two consecutive lines increases(SEM images in the left of figure 3) causing more energy distribution in higher order diffraction modes. This phenomenon can be explained on the basis of N slit experiment. A set of N parallel slits illuminated by a monochromatic wave show that the intensity of the light passing through the slits will depend upon the angle $\theta$ between the direction of the light propagation and the perpendicular to screen as[34];

$$I_\theta = I_0 \left[\frac{\sin\beta}{\beta}\right]^2 \left[\frac{\sin(N\alpha)}{\sin(\alpha)}\right]^2$$

$$\text{Where } \beta = (\pi/\lambda).b\sin\theta, \quad \alpha = (\pi/\lambda).d\sin\theta$$

$I_0$ is the intensity of the light in the centre of the diffraction for a single slit of width b, d is the distance between the slits and $k=2\pi/\lambda$ is the wave vector and $\lambda$ is the wavelength. The first term of the equation in the square brackets describes the Fraunhofer diffraction on one slit and the second term describes the interference from N point sources. The intensity of the light passing through the slit is proportional to b while the width of the diffraction pattern is proportional to 1/b. For this reason the intensity of the light $I_0$ in the centre of diffraction pattern will be proportional to $b^2$ causing the reduction in the intensity of $0^{th}$ order diffraction pattern.

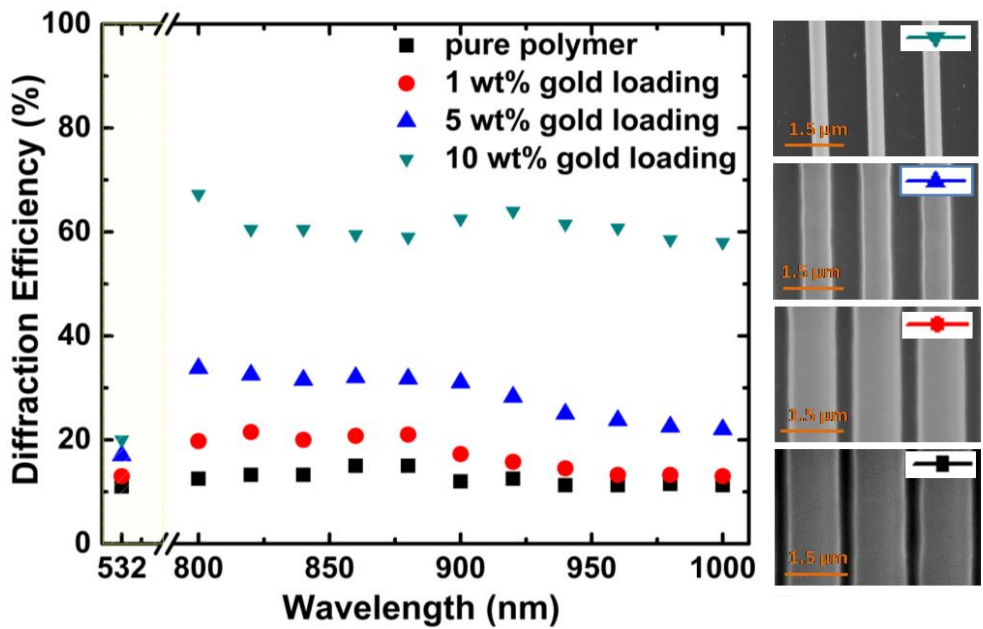

*Figure 3: Measured diffraction efficiency of the grating structures fabricated in 0wt%, 1wt%, 5wt% and 10Wt% gold loaded polymer matrix. Right panel are the SEM images of respective grating with periodicity of 1.5µm.*

The SEM images of large area gating structures with varied spacing have also been fabricated in 0 wt%, 1wt%, 5wt% and 10wt% gold loaded sartomer and are shown in figure S5, S6, S7, S8 (supporting information) respectively. Further to show the capability of writing functional 2D and 3D plasmonic devices we have successfully fabricated mesh structures with varied gold loading. We have fabricated 2D mesh structures in 0 wt%, 1wt%, 5wt% and 10wt% gold loaded sartomer (figure S9). Figure 4(a) is the SEM image of the 2D mesh structure written inside 10wt% gold loaded polymer. The diffraction pattern of red laser (712 nm wavelength ) for this 2D mesh structure showing bright and intense maxima can be seen in the inset of figure 4(a). A 3D mesh structure having 3 layers fabricated inside 1wt% gold loaded polymer is sown in figure 4(b). The SEM images of the 3 layer mesh structure at different magnifications are shown in figure S10.

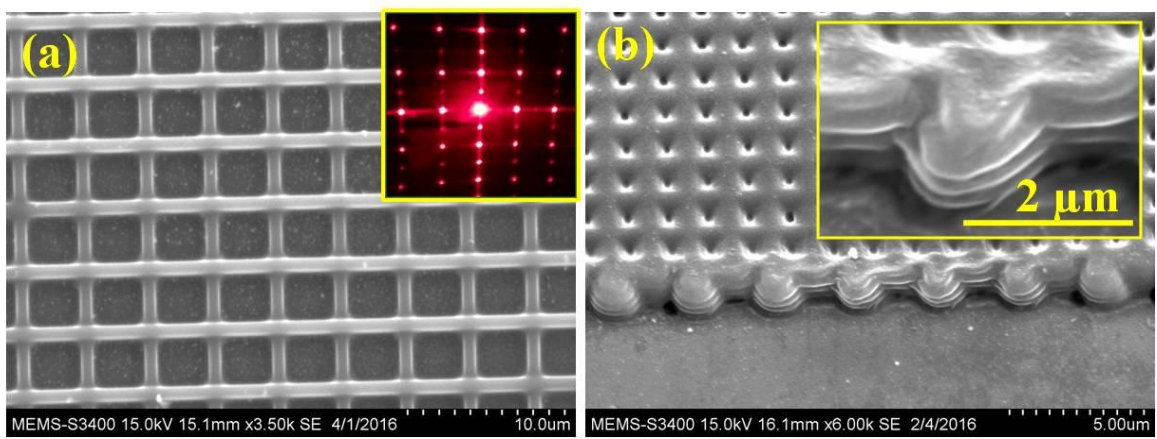

*Figure 4: (a) SEM image of the 2D mesh structure written inside 10wt% gold loaded polymer. Inset in 4(a) is the diffraction pattern of red laser (712 nm) from the respective 2D mesh structure. (b) 3 D (3 layer) mesh structure fabricated inside 1wt% gold loaded polymer*

Further to investigate the influence of laser power on feature sizes of 0wt, 1wt%, 5wt% and 10 wt% gold loaded polymeric lines, a systematic study was performed. Different line widths in the range of 390 nm to 750 nm were achieved by varying the concentration of gold precursor loading as shown in figure 5(a). SEM micrograph of 5% gold loaded polymeric structures in figure 5(b) shows feature size of ~ 390nm with gold nanoparticles embedded inside the polymer matrix. Gold nanoparticle loaded polymer lines can only be formed in the laser-exposed regions. The amount of free radicals generated by Lucirin TPOL is responsible for initiation of polymerization process of the sartomer as well as reduction of the gold precursor in the laser irradiated region.

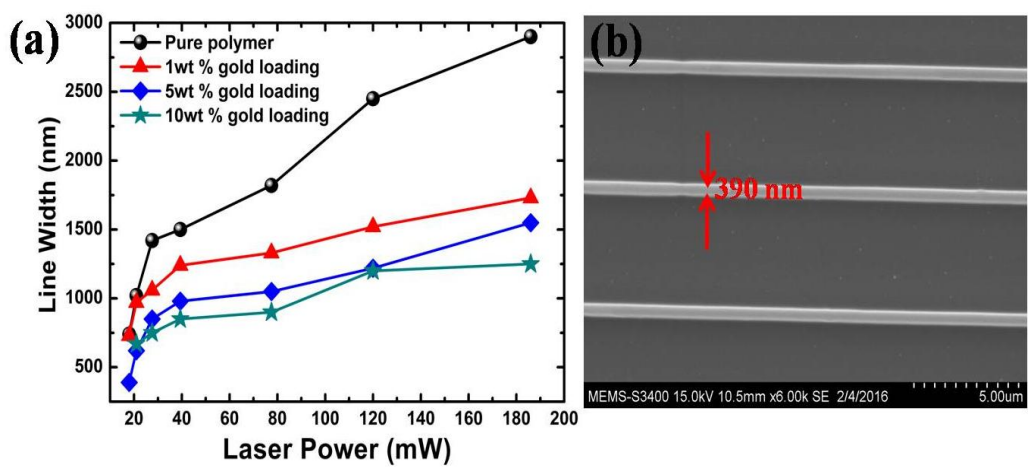

*Figure 5: (a) Plot of laser power vs. line width for pure polymer and gold loaded polymeric structures with different loading. (b) SEM image of 5wt% gold loaded polymeric lines showing sub-wavelength feature size of 390nm.*

The threshold value of the laser power for writing the structures within polymer with gold precursor up to 5wt% was found to be roughly the same (~18mW) as that for pure polymeric photoresist. The line width was found to increase with increasing laser power and decrease with increase in gold concentration up to nearly 5wt% loading at ~18mW. This decrease in line width with increasing gold concentration can be attributed to the coupling between free radical polymerization and gold reduction. It can be understood as the two processes compete against each other giving rise to finer features. It was observed that for 10wt% gold loaded samples the threshold value of laser power for writing increased to 24mW; additionally the feature size also increased as compared to 5wt% gold loaded polymeric lines. A minimum feature size of ~ 670nm was obtained for 10wt% gold loaded polymeric structures near threshold value, below which writing was not possible. This increase in feature size may be attributed to the increase in the laser threshold value. Initially there is a competition between oligomers and gold ions for free radical, which leads to reduction in line width up to 5wt% gold loading. It is well known that small gold nanoparticles are more mobile than the gold ions[19], which gets evident by the increased line width for 10wt% gold precursor loading.

Writing process starts with the absorption of 800nm light by the photo-initiator (Lucirin- TPOL). Free radical generated by TPA starts to reduce the gold precursor leading to the formation of gold nanoparticles in the polymeric matrix. Unlike SU8-gold precursor film, the sartomer-gold precursor film doesn't show any signs of degradation under normal room light or by simple mixing. The gold films were found to be stable even after a week of storage time. The detailed mechanism of simultaneous photo-polymerization and gold reduction is presented below[35]:

Two photon initiation of lucirin-TPOL gives rise to two free radicals

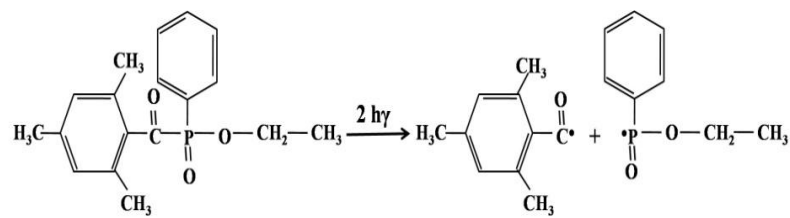

Free radical polymerization initiation of monomer 1:

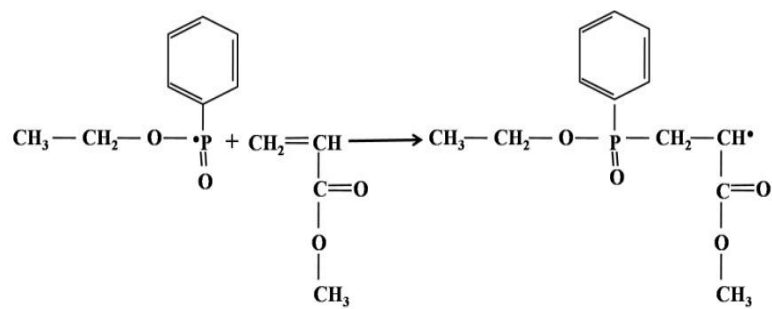

Propagation of chain polymerization reaction with monomer 2:

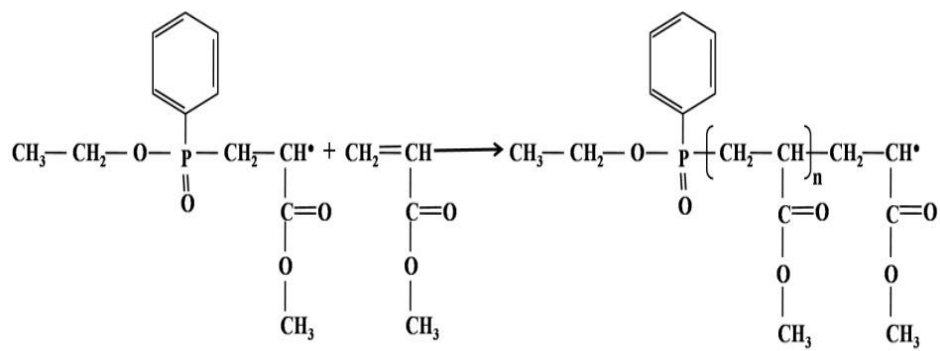

Termination by free radical combination

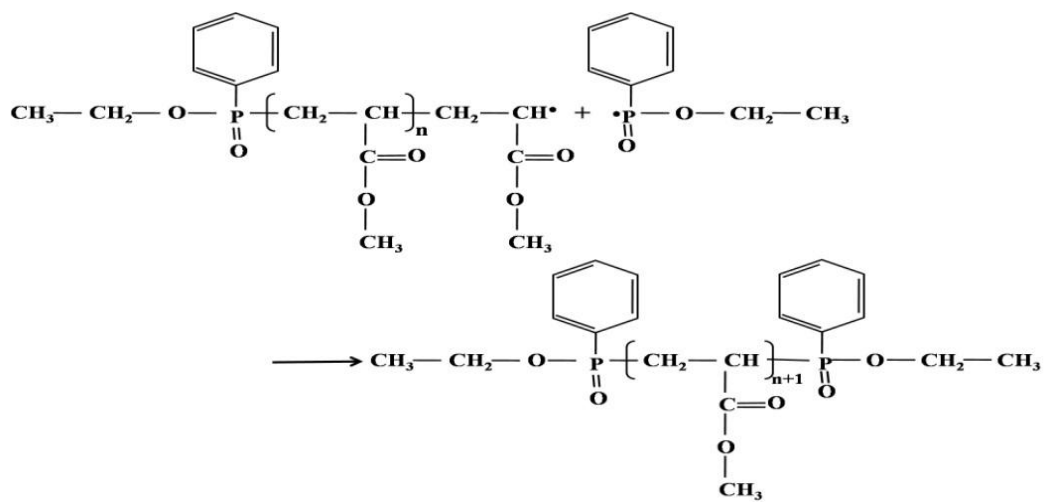

Reduction of gold from $Au^{+3}$ to $Au^{+2}$ by the free radical:

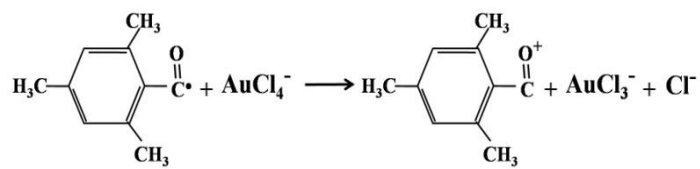

Fast disproportionation of $Au^{+2}$:

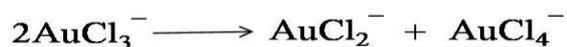

Disproportionation of $Au^{+1}$ and $Au^{+2}$ to reduce $Au^{+3}$ and $Au^0$:

$$AuCl_2^- + AuCl_3^- \longrightarrow AuCl_4^- + Au^0 + Cl^-$$

Gold nanoparticle formation

$$nAu^0 \longrightarrow Au_n$$

Figure 6(a) (red curve) shows the variation in the line width for various wt% gold loaded polymeric lines written at constant laser power. Blue curve measures the spacing between the two lines for various gold loading, written at constant periodicity of 1.5μm. Line span i.e maximum and minimum line widths obtained at a particular gold loading has also been plotted in figure 6(a). The writing threshold is higher for 10wt% gold loading with less spanning of line width. It is likely that sharing of photoinitiator between gold and oligomer causes this effect. It is apparent that with the increase in the gold loading, more and more hungry gold ions participate in the photoreduction process leading to less spanning in the line width. The upper threshold value for writing the structures with 1%, 5% and 10% loading is~186 mW beyond which the sample starts to ablate. Figure 6 (b) and (c) shows the grating structures written inside pure resin and 10wt% gold doped resin, respectively. The periodicity of both these gratings in this figure was kept constant to show the reduction in line width with increase in the gold loading at constant laser power.

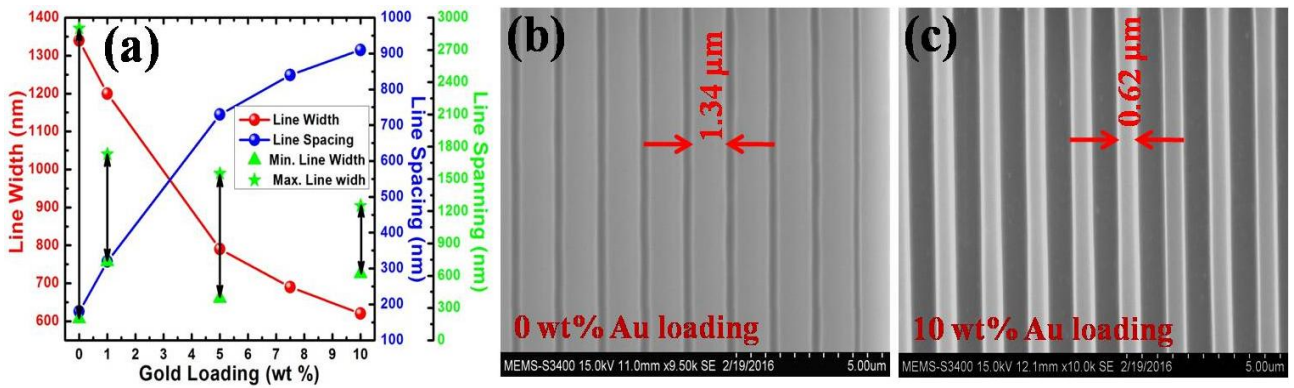

*Figure 6: (a) Measurement of line width for various wt% gold loaded polymeric lines (red cure). Blue curve represents the increase in line spacing with increase in gold loading measured at constant periodicity of 1.5μm for the gratings. Line spanning at various gold concentrations has also been plotted (green curve).SEM micrographs of grating structures fabricated in (b)0wt% and(c)10wt% gold loaded polymer matrix respectively at constant laser power just above the threshold value. The grating period is 1.5μm and constant for both the structures.*

**Conclusion:**

In conclusion, we have fabricated large area 2D and 3D functional subwavelength nanostructures in the form of diffraction gratings of gold loaded polymer composite by femto-second laser based two-photon lithography using single photon photoinitiators only in pure and metal precursor doped polymers. Free radicals generated by lucirin-TPOLare found to be responsible for this process which eradicates the need of an expensive two-photon dye. Two photon initiated photoreduction of gold precursor and simultaneous free radical polymerization of sartomer facilitates the fabrication of gold nanoparticle embedded 2D polymeric matrix. This phenomenon can be extended to the fabrication of 3D gold loaded polymeric structures. Features with line widths as small as 390nm has been observed for 5wt% gold doped resin without using any expensive two photon dye. The impact of critical process parameters such as laser power, gold doping etc on feature sizes has also been studied. This approach is expected to pave way for in-situ fabrication of high gold loaded polymer matrix which holds potential application towards novel 3D metamaterials, plasmonics and other optical technologies.

**Acknowledgements:**

We would like to acknowledge the generous support from DST CERI India for their support in establishing the experimental facility.

# References:

[1] W. U. Huynh, "Hybrid Nanorod-Polymer Solar Cells," *Science (80-. ).*, vol. 295, no. 5564, pp. 2425–2427, 2002.

[2] C. B. Fox, J. Kim, E. B. Schlesinger, H. D. Chirra, and T. A. Desai, "Fabrication of micropatterned polymeric nanowire arrays for high-resolution reagent localization and topographical cellular control," *Nano Lett.*, vol. 15, no. 3, pp. 1540–1546, 2015.

[3] D. . R. . Smith, J. B. Pendry, and M. . C. . K. . Wiltshire, "Metamaterials and Negative Refractive Index," *Science (80-. ).*, vol. 305, no. 5685, pp. 788–792, 2010.

[4] S. Shukla, K. T. Kim, A. Baev, Y. K. Yoon, N. M. Litchinitser, and P. N. Prasad, "Fabrication and characterization of gold-polymer nanocomposite plasmonic nanoarrays in a porous alumina template," *ACS Nano*, vol. 4, no. 4, pp. 2249–2255, 2010.

[5] E. P. Furlani and A. Baev, "Optical nanotrapping using cloaking metamaterial," *Phys. Rev. E - Stat. Nonlinear, Soft Matter Phys.*, vol. 79, no. 2, pp. 1–6, 2009.

[6] D. Strickland and G. Mourou, "Compression of Amplified Chirped Optical Pulses *," *Opt. Commun.*, vol. 56, no. 3, pp. 219–221, 1985.

[7] Y. M. Song, J. S. Yu, and Y. T. Lee, "Antireflective submicrometer gratings on thin-

film silicon solar cells for light-absorption enhancement.," *Opt. Lett.*, vol. 35, no. 3, pp. 276–278, 2010.

[8] L. Chen, J. J. Wang, F. Walters, X. Deng, M. Buonanno, S. Tai, and X. Liu, "Large flexible nanowire grid visible polarizer made by nanoimprint lithography," *Appl. Phys. Lett.*, vol. 90, no. 6, pp. 2005–2008, 2007.

[9] T. Weber, T. Käsebier, E.-B. Kley, and A. Tünnermann, "Broadband iridium wire grid polarizer for UV applications.," *Opt. Lett.*, vol. 36, no. 4, pp. 445–447, 2011.

[10] S. Maruo, O. Nakamura, and S. Kawata, "Three-dimensional microfabrication with two-photon-absorbed photopolymerization.," *Opt. Lett.*, vol. 22, no. 2, pp. 132–4, Jan. 1997.

[11] S. Kawata, H. B. Sun, T. Tanaka, and K. Takada, "Finer features for functional microdevices.," *Nature*, vol. 412, no. 6848, pp. 697–8, Aug. 2001.

[12] K. S. Lee, R. H. Kim, D. Y. Yang, and S. H. Park, "Advances in 3D nano/microfabrication using two-photon initiated polymerization," *Prog. Polym. Sci.*, vol. 33, no. 6, pp. 631–681, 2008.

[13] A. Biswas, I. S. Bayer, A. S. Biris, T. Wang, E. Dervishi, and F. Faupel, "Advances in top-down and bottom-up surface nanofabrication: techniques, applications & future prospects.," *Adv. Colloid Interface Sci.*, vol. 170, no. 1–2, pp. 2–27, Jan. 2012.

[14] C. N. LaFratta, J. T. Fourkas, T. Baldacchini, and R. A. Farrer, "Multiphoton fabrication," *Angew. Chemie - Int. Ed.*, vol. 46, no. 33, pp. 6238–6258, 2007.

[15] L. L. Erskine, A. a Heikal, S. M. Kuebler, M. Rumi, X. Wu, S. R. Marder, and J. W. Perry, "Two-photon polymerization initiators for three- dimensional optical data storage and microfabrication," *Solid State Phys.*, vol. 398, no. March, pp. 51–54, 1999.

[16] K. D. Belfield, X. Ren, E. W. Van Stryland, D. J. Hagan, V. Dubikovsky, E. J. Miesak, and R. V June, "Near-IR Two-Photon Photoinitiated Polymerization Using a Fluorone / Amine Initiating System," *J. Am. Chem. Soc*, vol. 122, no. 6, pp. 1217–1218, 2000.

[17] G. Witzgall, R. Vrijen, E. Yablonovitch, V. Doan, and B. J. Schwartz, "Single-shot two-photonexposure of commercial photoresist for the production of three-dimensionalstructures," *Opt. Lett.*, vol. 23, no. 22, pp. 1745–1747, 1998.

[18] T. Baldacchini, A.-C. Pons, J. Pons, C. N. LaFratta, J. T. Fourkas, Y. Sun, and M. J. Naughton, "Multiphoton laser direct writing of two-dimensional silver structures," *Opt. Express*, vol. 13, no. 4, p. 1275, 2005.

[19] S. Shukla, X. Vidal, E. P. Furlani, M. T. Swihart, K. T. Kim, Y. K. Yoon, A. Urbas, and P. N. Prasad, "Subwavelength direct laser patterning of conductive gold nanostructures by simultaneous photopolymerization and photoreduction," *ACS Nano*, vol. 5, no. 3, pp. 1947–1957, 2011.

[20] S. . Oldenburg, R. . Averitt, S. . Westcott, and N. . Halas, "Nanoengineering of optical resonances," *Chem. Phys. Lett.*, vol. 288, no. 2–4, pp. 243–247, 1998.

[21] M. S. Rill, C. Plet, M. Thiel, I. Staude, G. von Freymann, S. Linden, and M. Wegener, "Photonic metamaterials by direct laser writing and silver chemical vapour deposition.," *Nat. Mater.*, vol. 7, no. 7, pp. 543–6, Jul. 2008.

[22] S. Shukla, R. Kumar, A. Baev, A. S. L. Gomes, and P. N. Prasad, "Control of spontaneous emission of CdSe nanorods in a multirefringent triangular lattice photonic crystal," *J. Phys. Chem. Lett.*, vol. 1, no. 9, pp. 1437–1441, 2010.

[23] S. Shukla, A. Baev, H. Jee, R. Hu, R. Burzynski, Y. K. Yoon, and P. N. Prasad, "Large-area, near-infrared (IR) photonic crystals with colloidal gold nanoparticles embedding," *ACS Appl. Mater. Interfaces*, vol. 2, no. 4, pp. 1242–1246, 2010.

[24] J. Henzie, J. Lee, M. H. Lee, W. Hasan, and T. W. Odom, "Nanofabrication of plasmonic structures.," *Annu. Rev. Phys. Chem.*, vol. 60, pp. 147–65, 2009.

[25] K. Kaneko, H. B. Sun, X. M. Duan, and S. Kawata, "Two-photon photoreduction of metallic nanoparticle gratings in a polymer matrix," *Appl. Phys. Lett.*, vol. 83, no. 7, pp. 1426–1428, 2003.

[26] S. Shukla, E. P. Furlani, X. Vidal, M. T. Swihart, and P. N. Prasad, "Two-photon lithography of sub-wavelength metallic structures in a polymer matrix," *Adv. Mater.*, vol. 22, no. 33, pp. 3695–3699, 2010.

[27] J. F. Xing, X. Z. Dong, W. Q. Chen, X. M. Duan, N. Takeyasu, T. Tanaka, and S. Kawata, "Improving spatial resolution of two-photon microfabrication by using photoinitiator with high initiating efficiency," *Appl. Phys. Lett.*, vol. 90, no. 13, pp. 31–33, 2007.

[28] P. Tayalia, C. R. Mendonca, T. Baldacchini, D. J. Mooney, and E. Mazur, "3D cell-migration studies using two-photon engineered polymer scaffolds," *Adv. Mater.*, vol. 20, no. 23, pp. 4494–4498, 2008.

[29] V. Tribuzi, D. S. Corrêa, W. Avansi, C. Ribeiro, E. Longo, and C. R. Mendonça, "Indirect doping of microstructures fabricated by two-photon polymerization with gold nanoparticles.," *Opt. Express*, vol. 20, no. 19, pp. 21107–13, 2012.

[30] J.-F. Xing, M.-L. Zheng, and X.-M. Duan, "Two-photon polymerization microfabrication of hydrogels: an advanced 3D printing technology for tissue engineering and drug delivery.," *Chem. Soc. Rev.*, vol. 44, pp. 5031–5039, 2015.

[31] X. Zhou, Y. Hou, and J. Lin, "A review on the processing accuracy of two-photon polymerization," *AIP Adv.*, vol. 5, no. 3, 2015.

[32] T. Baldacchini, C. N. LaFratta, R. A. Farrer, M. C. Teich, B. E. A. Saleh, M. J. Naughton, and J. T. Fourkas, "Acrylic-based resin with favorable properties for three-dimensional two-photon polymerization," *J. Appl. Phys.*, vol. 95, no. 11 I, pp. 6072–6076, 2004.

[33] K. K. Seet, S. Juodkazis, V. Jarutis, and H. Misawa, "Feature-size reduction of photopolymerized structures by femtosecond optical curing of SU-8," *Appl. Phys. Lett.*, vol. 89, no. 2, pp. 10–13, 2006.

[34] E. Hecht, *Optics*. Addison-Wesley, 2002.

[35] R. E. Medsker, M. Chumacero, E. R. Santee, A. Sebenik, and H. J. Harwood, "31P-NMR characterization of chain ends in polymers and copolymers prepared using Lucirin TPO as a photoinitiator," *Acta Chim. Slov.*, vol. 45, no. 4, pp. 371–388, 1998.